\documentclass[aps,prb,twocolumn,showpacs,floatfix,superscriptaddress,unsortedaddress]{revtex4-1}
\usepackage{amsmath}
\usepackage{amsfonts}
\usepackage{amssymb}
\usepackage{euscript}
\usepackage{bm}
\usepackage{epsfig}
\usepackage{graphicx}
\newcommand{\RNum}[1]{\uppercase\expandafter{\romannumeral #1\relax}}
\providecommand{\U}[1]{\protect \rule{.1in}{.1in}}

\usepackage[usenames,dvipsnames]{color}
 \usepackage[colorlinks=true, urlcolor=blue,linkcolor=Violet,citecolor=PineGreen, pdfborder={0 0 0}]{hyperref}
\usepackage{cleveref}
\usepackage{placeins}
 \crefname{equation}{Eq.~}{Eqs.~}
 \crefname{figure}{Fig.}{Figs.}
  \crefname{section}{Sec.}{Sec.}

\begin{document}

 \title{Diagrammatic Monte Carlo study of the acoustic and the BEC polaron}
\author{Jonas Vlietinck}
\affiliation{Department of Physics and Astronomy, Ghent University,
 Proeftuinstraat 86, 9000 Gent, Belgium}
\author{Wim Casteels}
\affiliation{TQC, Universiteit Antwerpen, Universiteitsplein 1, 2610 Wilrijk, Belgium} 
\author{Kris Van Houcke}
\affiliation{Department of Physics and Astronomy, Ghent University,
 Proeftuinstraat 86, 9000 Gent, Belgium}
\affiliation{Laboratoire de Physique Statistique, Ecole Normale Sup{\' e}rieure, UPMC, Universit{\' e} Paris Diderot,
 CNRS, 24 rue Lhomond, 75231 Paris Cedex 05, France}
\author{Jacques Tempere}
\affiliation{TQC, Universiteit Antwerpen, Universiteitsplein 1, 2610 Wilrijk, Belgium} 
\affiliation{Lyman Laboratory of Physics, Harvard University,
Cambridge, Massachusetts 02138, USA}
\author{Jan Ryckebusch}
\affiliation{Department of Physics and Astronomy, Ghent University,
 Proeftuinstraat 86, 9000 Gent, Belgium}
 \author{Jozef T. Devreese}
\affiliation{TQC, Universiteit Antwerpen, Universiteitsplein 1, 2610 Wilrijk, Belgium}

\begin{abstract}
We consider two large polaron systems that are described by a
Fr\"{o}hlich type of Hamiltonian, namely the Bose-Einstein condensate
(BEC) polaron in the continuum and the acoustic polaron in a
solid.  We present ground-state energies of these two systems
calculated with the Diagrammatic Monte Carlo (DiagMC) method and with
a Feynman all-coupling approach. The DiagMC method evaluates up to
very high order a diagrammatic series for the polaron Green's
function. The Feynman all-coupling approach is a variational method
that has been used for a wide range of polaronic problems. For the
acoustic and BEC polaron both methods provide remarkably similar
non-renormalized ground-state energies that are obtained after
introducing a finite momentum cutoff. For the renormalized
ground-state energies of the BEC polaron, there are relatively large
discrepancies between the DiagMC and the Feynman predictions.  These
differences can be attributed to the renormalization procedure for the
contact interaction.
\end{abstract}
\pacs{71.38.Fp,02.70.Ss,67.85.Bc}
\maketitle

\section{Introduction}\label{intro}
By virtue of the Coulomb interaction the presence of a charge carrier
in a charged lattice induces a polarization. This effect is well-known
from the description of an electron or a hole in a polar or ionic
semiconductor. The term polaron was coined by Landau in 1933
\cite{Landau} to denote the quasiparticle comprised of a charged
particle coupled to a surrounding polarized lattice. For
lattice-deformation sizes of the order of the lattice parameter, one
refers to the system as a small or Holstein polaron
\cite{PhysRevB.53.9666,marchand}. For lattice-deformation sizes that
are large compared to the lattice parameter, the lattice can be
treated as a continuum. This system is known as a large polaron for
which Fr\"{o}hlich proposed the model Hamiltonian~\cite{Frohlish}
\begin{equation}
\begin{split}
\hat{H}_{pol}  & =\sum_{\mathbf{k}}\frac{\hbar^{2}\mathbf{k}^{2}}{2m}\hat{c}%
_{\mathbf{k}}^{\dag}\hat{c}^{\phantom{\dag}}_{\mathbf{k}}+\sum_{\mathbf{k}}\hbar \omega(\mathbf{k}%
)\hat{b}_{\mathbf{k}}^{\dag}\hat{b}^{\phantom{\dag}}_{\mathbf{k}} \\
& +\sum_{\mathbf{k},\mathbf{q}}V (\mathbf{q} ) \hat{c}_{\mathbf
{k}+\mathbf{q}}^{\dag}\hat{c}^{\phantom{\dag}}_{\mathbf{k}}\left(  \hat{b}_{-\mathbf{q}}^{\dag
}+\hat{b}^{\phantom{\dag}}_{\mathbf{q}}\right)  .
\end{split}
\label{PolHam}
\end{equation}
Here, the $\hat{c}_{\mathbf{k}}^{\dag}$ ($\hat{c}^{\phantom{\dag}}_{\mathbf{k}}$) are
the creation (annihilation) operators of the charge carriers with band
mass $m$ and momentum $\mathbf{k}$. The second term in the above
Hamiltonian gives the energy of the phonons which carry the
polarization. Thereby, the operator $\hat{b}_{\mathbf{k} }^{\dag}$
($\hat{b}^{\phantom{\dag}}_{\mathbf{k}}$) creates (annihilates) a phonon with wave
vector $\mathbf{k}$ and energy $\hbar \omega(\mathbf{k})$.  The last
term in \cref{PolHam} denotes the interaction between the charge
carrier and the phonons. A plethora of physical phenomena can be
described by the above Fr\"{o}hlich type of Hamiltonian by varying the
dispersion $\omega(\mathbf{k})$ and the interaction strength $V
(\mathbf{q} )$.
Fr\"{o}hlich considered the special situation of  longitudinal optical (LO) phonons which
are dispersionless $\omega(\mathbf{k})=\omega_{LO}$.  In the LO limit,
the interaction amplitude $V(\mathbf{q})$ in Eq.~(\ref{PolHam}) adopts
the form
\begin{equation}
V _{LO}(\mathbf{q})=-i\frac{\hbar \omega_{LO}}{q}\left(  
\frac{4\pi \alpha_{LO}}{\mathcal{V}} \right)
^{1/2}\left(  \frac{\hbar}{2m\omega_{LO}}\right)  ^{1/4}.
\label{eq:vlo}
\end{equation}
Here, $\mathcal{V}$ is the volume of the crystal and $\alpha_{LO}$ the dimensionless
coupling parameter:
\begin{equation}
\alpha_{LO}=\frac{e^{2}}{\hbar}\sqrt{\frac{m}{2\hbar \omega_{LO}}}\left(  \frac
{1}{\varepsilon_{\infty}}-\frac{1}{\varepsilon_{0}}\right)  \; ,
\label{eq:LOcouplingparameter}
\end{equation}
with $\varepsilon_{\infty}$ ($\varepsilon_{0}$) the electronic
(static) dielectric constants of the crystal and $e$ the charge of the
electron. The Fr\"{o}hlich polaron which is defined by the
Eqs.~(\ref{PolHam})-(\ref{eq:vlo}) and the dispersion
$\omega(\mathbf{k})=\omega_{LO}$, has no analytical solution. 

More generally, solutions to the Eq.~(\ref{PolHam}) describe a
quasiparticle interacting with a bath of non-interacting bosons 
 with energies $ \hbar \omega(\mathbf{k})$ through the
mediation of the interaction $V(\mathbf{q})$. One example is the
acoustic polaron which corresponds to the interaction of a charge
carrier with acoustic phonons \cite{PhysRevB.32.3515}. Another example
is the BEC polaron consisting of an impurity atom interacting with the
Bogoliubov excitations of an atomic Bose-Einstein condensate (BEC)
\cite{PhysRevLett.96.210401, PhysRevA.73.063604,
  PhysRevB.80.184504}. Other examples are an electron on a helium film
(``ripplopolaron'') \cite{ripplopolaron,PhysRevB.24.499,
  PhysRevB.39.4133} and a charge carrier in a piezoelectric
semiconductor (``piezopolaron'') \cite{PhysRevLett.12.241}.

Due to the relative simplicity of the model Hamiltonian of
\cref{PolHam} it is an ideal testing ground for conducting comparative
studies with various many-body techniques (see for example
Refs.~\cite{BoekDevreese, 2010arXiv1012.4576D} for an overview). The
weak coupling regime (small $\alpha_{LO}$) was described by
Fr\"{o}hlich with second-order perturbation theory \cite{Frohlish}
which is equivalent to the Lee-Low-Pines scheme using a canonical
transformation \cite{PhysRev.90.297}.  For the strong coupling regime
(large $\alpha_{LO}$) Landau and Pekar developed a variational
technique which predicts the formation of a bound state of the charge
carrier in his self-induced potential \cite{LandauPekar,
  Pekar}. Feynman developed a superior all-coupling approach
\cite{PhysRev.97.660, Feynman} which captures all the coupling
regimes. 

 A numerical solution of the Fr\"{o}hlich Hamiltonian of
Eq.~(\ref{PolHam}) with the interaction of Eq.~(\ref{eq:vlo}) has been
proposed in Refs.~\cite{Prokofev1998,mischenko2000B}. Thereby, a series
expansion for the polaron Green's function was evaluated with the aid
of a Diagrammatic Monte Carlo (DiagMC) method. The method is ``exact''
in the sense that the series expansion is convergent and sign-definite
and therefore it can be stochastically evaluated with a controllable error. The polaron's
energy is extracted from the asymptotic behavior of its Green's
function.

\begin{figure}[]
\includegraphics[width=\columnwidth] {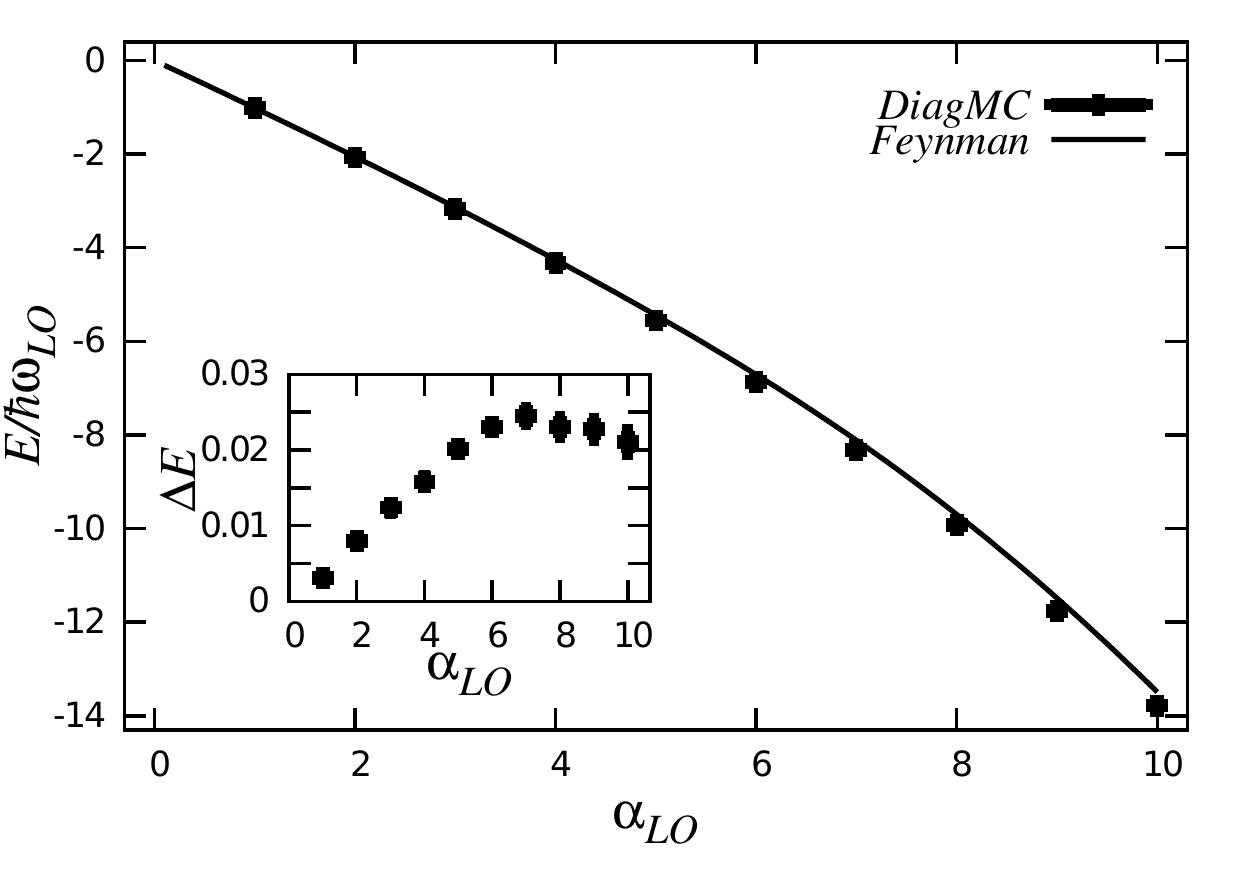}
\caption{Ground-state energies for the Fr\"{o}hlich polaron are shown as a
  function of the coupling strength $\alpha_{LO}$ of
  Eq.~(\ref{eq:LOcouplingparameter}). The inset shows the relative
  difference $\Delta E=\frac{E^{MC}-E^F}{E^{MC}}$, with $E^F$ ($E^{MC}$) the
  computed energy from the Feynman (DiagMC) approach.}
\label{fig:EpolEl} 
\end{figure}

Polaron systems are ideal for comparative studies of many-body
techniques. Examples of such studies for the Fermi polaron are
reported in Refs.~\cite{pietro,fermipol3D,fermipol2D}.  For the Fermi
polaron, a comparison has been made between the DiagMC method and the
variational technique which includes a limited number of particle-hole
excitations. It was demonstrated that a variational one particle-hole
calculation is already a good approximation, even for strong
interactions between the impurity and the particles in the Fermi sea
\cite{fermipol3D,fermipol2D}.  Recently a comparative study of the
neutron polaron has been conducted with quantum Monte Carlo and
effective field theories \cite{neutronpol}.  For the ground-state
energy of the Fr\"{o}hlich polaron of Eqs.~(\ref{PolHam}) and
(\ref{eq:vlo}) it has been shown in
Ref.~\cite{Prokofev1998} that
Feynman's approach reproduces the DiagMC results to a remarkable
accuracy.  We have reproduced those numerical results. As can be
appreciated from \cref{fig:EpolEl} the deviations between the
variational Feynman and DiagMC predictions for the ground-state
energies of the Fr\"{o}hlich polaron, are of the order of a few
percent, even for the large coupling strengths.
It is not clear, however, how accurate the Feynman technique is for
polaron systems described by a Hamiltonian of the type of
Eq.~(\ref{PolHam}) with alternate dispersions $\omega(\mathbf {k})$
and interaction amplitudes $V(\mathbf{q})$.  Indeed, Feynman's
approach is based on a variational action functional that models the
coupling to the phonons by a single phononic degree of freedom with a
variationally determined mass and harmonic coupling to the
electron. This is a rather natural choice for LO phonons, which are
dispersionless. However, it seems intuitively less suitable in
situations that the phonons' energies cover a finite range of
values. Thornber \cite{thornber}  has argued that in those situations,
Feynman's model is unlikely to yield accurate results for the system's
dynamical properties, but that the system's ground-state energy can
still be captured accurately. To our knowledge, this assertion has not
yet been sufficiently confirmed.  In order to remedy this situation,
in this work we compare polaron ground-state energies calculated with
the Feynman variational approach against DiagMC results. This will
allow us to test the robustness of the Feynman approach.  The two
prototypical polaron problems considered in this work are the BEC
polaron and the acoustic polaron. These problems have been selected
because they highlight complementary aspects. The effect of broadening
the range of phonon energies is captured by the acoustic polaron. The
BEC polaron problem allows one to additionally cover the
issues related to renormalizing $V(\mathbf{q})$.

The structure of this manuscript is as follows. In \cref{polmod} the
Hamiltonians for the BEC and acoustic polaron are introduced. In
Sects.~\ref{feyn} and \ref{diagmc} the adopted many-body methods for
obtaining the ground-state energies of those Hamiltonians are
sketched.  Results of the two techniques for the ground-state energies
of the BEC and acoustic polaron are contained in Sec.~\ref{result}.

\section{Large polaron models}\label{polmod}
\subsection{BEC polaron}\label{BEC}
The Hamiltonian of an impurity immersed in a bath of interacting
bosons \cite{PhysRevB.80.184504} is given by a sum of two terms $\hat{H}  = \hat{H}_B +
\hat{H}_I$ with,
\begin{equation}
 \begin{split}
\hat{H}_B  & =           \sum_{\mathbf{k}}  \epsilon_{\mathbf{k}}  ~ \hat{a}^{\dagger}_{\mathbf{k}}
\hat{a}^{\phantom{\dagger}}_{\mathbf{k}}
  +     \frac{1}{2\mathcal{V}} \sum_{\mathbf{k}, \mathbf{k}', \mathbf{q}}
V_{BB}(\mathbf{q})  ~
\hat{a}^{\dagger}_{\mathbf{k}'-\mathbf{q} } \hat{a}^{\dagger}_{\mathbf{k}+\mathbf{q}}
\hat{a}^{\phantom{\dagger}}_{\mathbf{k} } \hat{a}^{\phantom{\dagger}}_{\mathbf{k}' } \; , \\
\hat{H}_I  & =         \sum_{\mathbf{k}}  \frac{\hbar^2\mathbf{k}^2}{2m_I}  ~ \hat{c}^{\dagger}_{\mathbf{k}}
\hat{c}^{\phantom{\dagger}}_{\mathbf{k}}
 + \frac{1}{\mathcal{V}} \sum_{\mathbf{k}, \mathbf{k}', \mathbf{q}}
V_{IB}(\mathbf{q})  ~
\hat{c}^{\dagger}_{\mathbf{k}+\mathbf{q} } \hat{c}^{\phantom{\dagger}}_{\mathbf{k}}
\hat{a}^{\dagger}_{\mathbf{k}'-\mathbf{q} } \hat{a}^{\phantom{\dagger}}_{\mathbf{k}' } 
\; .  \\
\label{eq:ham}
\end{split}
\end{equation}
The operators $ \hat{a}^{\dagger}_{\mathbf{k}} ( \hat{a}^{\phantom{\dagger}}_{\mathbf{k}})$ create (annihilate) bosons
with momentum $\mathbf{k}$, mass $m$ and energy $\epsilon_{\mathbf{k}} =\hbar^2 \mathbf{k}^2 / 2 m$. Further, 
$\mathcal{V}$
is the volume of the system.  
The operators $ \hat{c}^{\dagger}_{\mathbf{k}} (\hat{c}^{\phantom{\dagger}}_{\mathbf{k}})$ create (annihilate) the
impurity with momentum $\mathbf{k}$ and mass $m_I$. 
The boson-boson and impurity-boson interactions in momentum space are  
$V_{BB}(\mathbf{q})$ and $V_{IB}(\mathbf{q})$.
These potentials are replaced by the
pseudopotentials $g_{BB}$ and $g_{IB}$. These constants are chosen
such that the two-body scattering properties in vacuum are correctly
reproduced.  The sum of all vacuum ladder diagrams, given by the
$T$-matrix, represents all possible ways in which two particles can
scatter in vacuum. For zero momentum and frequency the $T$-matrix is
given by $T(0)$:
\begin{equation}
 T(0)=g_{IB}-g_{IB} \sum_k\frac{2 m_r}{\hbar^2k^2} T(0) \;,
 \label{eq:lad}
\end{equation}
with $m_r = (1/m_I+1/m)^{-1}$ the reduced mass. For low-energy collisions the first-order Born approximation can be
applied to model the boson-boson and boson-impurity collisions. As a result, $g_{IB}=\frac{2 \pi a_{IB} \hbar^2}{m_r}$,
with $a_{IB}$
the boson-impurity scattering length and  $g_{BB}=\frac{4 \pi a_{BB} \hbar^2}{m}$, with $a_{BB}$
the boson-boson scattering length.

In the Bogoliubov approximation \cite{Pitaevskii}, the Hamiltonian
$\hat{H}_B$ of \cref{eq:ham} is written in the diagonal form
\begin{equation}
\hat{H}_B \approx E_0 + \sum_{\mathbf{k}\neq 0} \hbar\omega(\mathbf{k}) \hat{b}^{\dagger}_{\mathbf{k}}
\hat{b}^{\phantom{\dagger}}_{\mathbf{k}} \; ,
\end{equation}
where the operators $\hat{b}^{\dagger}_{\mathbf{k}} (\hat{b}^{\phantom{\dagger}}_{\mathbf{k}})$ create (annihilate)
Bogoliubov quasi-particles.
The quasi-particle vacuum energy is
\begin{equation}
E_0 = \frac{\mathcal{V}}{2} n^2 g_{BB} + \frac{1}{2} \sum_{\mathbf{k}\neq 0} \bigg(  \hbar\omega(\mathbf{k}) -
\epsilon_{\mathbf{k}} -   n_0 g_{BB}   \bigg)  \; ,
\end{equation}
with $n=N/\mathcal{V}$ the total density and $n_0=N_0/\mathcal{V}$ the
density of the condensed bosons.  The average total particle number
$N=\langle\hat{N}\rangle$ is fixed, with 
\begin{equation}
\hat{N} = N_0 + \sum_{\mathbf{k}\neq 0}  \hat{a}^{\dagger}_{\mathbf{k}} \hat{a}^{\phantom{\dagger}}_{\mathbf{k}} \; ,
\end{equation}
and $N_0$ the number of bosons in the condensate. 
The collective Bogoliubov excitations obey the dispersion relation 
\begin{equation}
\hbar \omega(\mathbf{k}) = \sqrt{(\epsilon_{\mathbf{k}}+n_0g_{BB})^2 - (n_0g_{BB})^2} \; .
\end{equation}
At long wavelengths, the spectrum becomes $\omega(\mathbf{k}) =
|\mathbf{k}| c$, which is characteristic of a sound wave with velocity
$c=\sqrt{n_0g_{BB}/m}$.  The excitation spectrum is conveniently
written in the form
\begin{equation}
\omega(\mathbf{k}) = kc\sqrt{1+ \frac{(k\xi)^2}{2}} \; ,
\end{equation}
with $k=|\mathbf{k}|$ and $\xi=1/\sqrt{2mn_0g_{BB}}$ the healing
length of the Bose condensate.

Application of the Bogoliubov transformation to the impurity part $\hat{H}_{I}$ of \cref{eq:ham} gives \cite{PhysRevLett.96.210401,
PhysRevA.73.063604, PhysRevB.80.184504}
\begin{eqnarray}
\hat{H}_I & \approx &   \sum_{\mathbf{k}}  \frac{\hbar^2\mathbf{k}^2}{2m_I}  ~ \hat{c}^{\dagger}_{\mathbf{k}}
\hat{c}^{\phantom{\dagger}}_{\mathbf{k}} + n_0 g_{IB}
\nonumber \\
& + & \sum_{\mathbf{q}\neq0, \mathbf{k}} V_{BP}({\mathbf{q}})~  \hat{c}^{\dagger}_{\mathbf{k}+\mathbf{q} }
\hat{c}^{\phantom{\dagger}}_{\mathbf{k}} 
\big( \hat{b}^{\dagger}_{-\mathbf{q}} + \hat{b}^{\phantom{\dagger}}_{\mathbf{q}}\big)
\; ,
\end{eqnarray}
in which we have defined 
\begin{eqnarray}
V_{BP}({\mathbf{q}}) & = &  \frac{g_{IB}}{\mathcal{V}} \sqrt{\frac{N_0 \epsilon_{\mathbf{q}}}{\omega(\mathbf{q})}}\nonumber 
\\
& = &     \frac{g_{IB} \sqrt{N_0}}{\mathcal{V}} \bigg(  \frac{(\xi q)^2}{(\xi q)^2+2} \bigg)^{1/4}  \; .
\end{eqnarray}
For $g_{IB}=\frac{2 \pi a_{IB}\hbar^2}{m_r}$  a dimensionless coupling constant $\alpha_{IB}$ can be defined
\cite{PhysRevB.80.184504} 
\begin{equation}
\alpha_{IB}=\frac{a_{IB}^{2}}{a_{BB}\xi} \;.
\label{AlfBEC}
\end{equation}
The final expression for the Hamiltonian for the BEC polaron is given by
\begin{equation}
\begin{split}
\hat{H}_{BP} &=  E_0  + n_0 g_{IB} 
+ \sum_{\mathbf{k}}  \frac{\hbar^2\mathbf{k}^2}{2m_I}  ~ \hat{c}^{\dagger}_{\mathbf{k}}
\hat{c}^{\phantom{\dagger}}_{\mathbf{k}}
+ \sum_{\mathbf{k}\neq 0} \hbar\omega(\mathbf{k}) \hat{b}^{\dagger}_{\mathbf{k}}
\hat{b}^{\phantom{\dagger}}_{\mathbf{k}}
\\ &+ \sum_{\mathbf{q}\neq0, \mathbf{k}} V _{BP} ({\mathbf{q}})~  \hat{c}^{\dagger}_{\mathbf{k}+\mathbf{q} }
\hat{c}^{\phantom{\dagger}}_{\mathbf{k}} 
\big( \hat{b}^{\dagger}_{-\mathbf{q}} + \hat{b}^{\phantom{\dagger}}_{\mathbf{q}}\big) \; .
\end{split}
\label{eq:mapHam}
\end{equation}
Obviously, the $\hat{H}_{BP}$ has the format of a Fr\"{o}hlich-type of
Hamiltonian defined in Eq.~(\ref{PolHam}).  When presenting numerical results
for the BEC polaron, lengths will be expressed in units of $\xi$,
energies in units of $\frac{\hbar^2}{m \xi^2}$ and phonon wave
vectors in units of $1/ \xi$. In this way, all quoted variables are
dimensionless.  In the numerical calculations, we consider an $^{6}$Li
impurity in a Na condensate for which $m_{I}/m_{B}=0.263158$
\cite{PhysRevB.80.184504}.

\subsection{Acoustic polaron}
\label{ac}
 
In a crystal with two or more atoms per primitive cell, the
dispersion relation $\omega(\mathbf{k})$ for the phonons develops 
acoustic as well as optical branches.  The acoustic polaron comprises a
charge carrier interacting with the longitudinal acoustic phonons and
can be described by the Fr\"{o}hlich type of Hamiltonian of Eq.~(\ref{PolHam})
with the dispersion $\omega(\mathbf{k})=sk$,
with $s$ the sound velocity \cite{PhysRevB.32.3515}. For the acoustic polaron, the interaction
$V _{AC} (\mathbf{q})$ in the Fr\"{o}hlich Hamilonian adopts the form
\cite{PhysRevB.32.3515}:
\begin{equation}
V _{AC} (\mathbf{q})=\left(  \frac{4\pi \alpha_{AC}}{\mathcal{V}}\right)  ^{1/2}\frac{\hbar^{2}}%
{m}\sqrt{q} \;, 
\end{equation}
with $\mathcal{V}$ the volume of the crystal and $\alpha_{AC}$ a
dimensionless coupling parameter. When discussing results concerning
the acoustic polaron, lengths will be expressed in units of $\hbar /
(ms)$, energies in units of $ms^2$ and phonon wave vectors in units of
$ms/\hbar$.  The summations over the phonon momenta $\mid \mathbf{k}
\mid$ have a natural cut-off at the boundary $k_{0}$ of the first
Brillouin zone.  At strong coupling, the Feynman approach to the
acoustic polaron predicts the emergence of a self-induced binding
potential for the impurity (``self-trapped state''). For a system with
both Fr\"{o}hlich and acoustic phonons, the Feynman approach predicts
that the dominant mechanism for this transition is the interaction
with the acoustic phonons \cite{doi:10.1143/JPSJ.35.137}. Only
considering the acoustic phonons results in a transition of the first
order for $k_{0}>18$ and a critical point at $k_{0}\approx18$ and
$\alpha_{AC} \approx0.151$ \cite{PhysRevB.32.3515}. This transition
was also predicted by the path integral Monte Carlo method
\cite{fantoni}.

\section{Numerical methods}\label{Num}
\subsection{Feynman variational path integral}\label{feyn}

The Feynman approach is based on the Jensen-Feynman inequality for the free
energy $\mathcal{F}$ of a system with action $\mathcal{S}$ \cite{Feynman}:
\begin{equation}
\mathcal{F}\leq \mathcal{F}_{0}+\frac{1}{\hbar \beta}\left \langle \mathcal{S}%
-\mathcal{S}_{0}\right \rangle _{\mathcal{S}_{0}}\;.\label{JF}%
\end{equation}
Here, $\mathcal{F}_{0}$ is the free energy of a trial system with
action $\mathcal{S}_{0}$, $\left \langle ...\right \rangle
_{\mathcal{S}_{0}}$ denotes the expectation value with respect to the
trial system and $\beta = \left( k_{B}T\right) ^{-1}$ is the inverse
temperature. Feynman proposed a variational trial system of a charge
carrier harmonically coupled with spring frequency $W$ to a fictitious
particle with mass $M$. For $T=0$ the Jensen-Feynman inequality of
\cref{JF} applied to this system produces an upper bound $E_p^F$ for the
polaronic ground-state energy  \cite{PhysRev.97.660, Feynman}:
\begin{equation}
\begin{split}
E_{p}  & \leq \frac{3\hbar \Omega}{4}\frac{\left(  \sqrt{\left(  1+M/m_{I}%
\right)  }-1\right)  ^{2}}{1+M/m_{I}}\\
& +\sum_{\mathbf{k}}\frac{\left \vert V_{\mathbf{k}}\right \vert ^{2}}{\hbar}\int
_{0}^{\infty}du\mathcal{D}\left(  \mathbf{k},u\right)  \mathcal{M}\left(  \mathbf
{k},u\right) \;,
\end{split}
\label{UppBound}
\end{equation}
with $\Omega=W\sqrt{1+M/m\text{ }}$. The function $\mathcal{D}\left(  \mathbf
{k},u\right)  $ is the phonon Green's function in momentum-imaginary-time representation
$(\mathbf{k},\tau)$ %
\begin{equation}
\mathcal{D}\left(  \mathbf{k},\tau \right)  =-\theta \left(  \tau \right)
\exp \left[  -\omega(\mathbf{k})\tau \right]  \; ,
\label{eq:phononpropagator}
\end{equation}
where $\theta(\tau)$ is the Heaviside function. The memory function  $\mathcal{M}\left(  \mathbf{k},u\right)  $ is:%
\begin{eqnarray}
\mathcal{M}\left(  \mathbf{k},u\right)  & = & 
\exp \left[  -\frac{\hbar k^{2}}{2\left(
m_{I}+M\right)  } \right.
\nonumber \\
& & \times \left.\left(  u+\frac{M}{m_{I}}\frac{1-\exp \left[  -\Omega
u\right]  }{\Omega}\right)  \right] \; .
\end{eqnarray}
The $u$-integral in \cref{UppBound} is of the following form:%
\begin{equation}
\int_{0}^{\infty}du\exp \left[  -au+be^{-u}\right]  =-\left(  -b\right)
^{-a}\Gamma \left(  a,-b,0\right)  \;,
\end{equation}
with $\Gamma \left( a,z_{0},z_{1}\right) =
\int_{z_{0}}^{z_{1}}t^{a-1}e^{-t}dt $ the generalized incomplete gamma
function.  The parameters $M$ and $\Omega$ are used to minimize the
upper bound for the ground state energy of \cref{UppBound}. This
approach captures the different coupling regimes.

\subsection{One-body propagator and DiagMC}\label{diagmc}

\begin{figure}[h]
\includegraphics[width=\columnwidth] {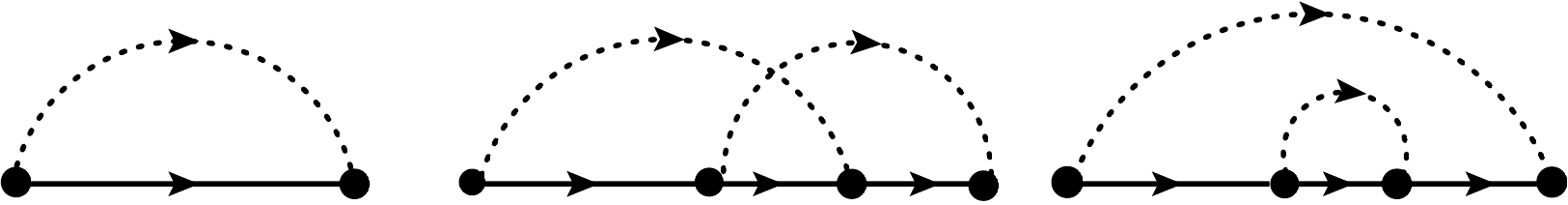}
\caption{Irreducible diagrams for the polaron's self-energy $\Sigma(\tau)$.  Imaginary
  time runs from left to right. A solid line represents a
  free-impurity propagator and a dashed line stands for an elementary
  excitation. The interaction vertices are denoted by dots.}
\label{fig:irrdiag}
\end{figure}

The Green's function of the polaron in the $(\mathbf{k},\tau)$ representation is defined as:
\begin{equation}
G(\mathbf{k},\tau)=-\theta(\tau)\langle \mathrm{vac}|\hat{c}_{\mathbf{k}%
}^{\phantom{\dagger}}(\tau)\hat{c}_{\mathbf{k}}^{\dagger}(0)|\mathrm{vac}%
\rangle,
\label{eq:Greenstime}
\end{equation}
with
\begin{equation}
\hat{c}_{\mathbf{k}}^{\phantom{\dagger}}(\tau)=e^{\hat{H}\tau}\hat
{c}_{\mathbf{k}}^{\phantom{\dagger}}e^{-\hat{H}\tau},
\end{equation}
the annihilation operator in the Heisenberg representation and $|\mathrm{vac}%
\rangle$ the vacuum state. The BEC polaron Hamiltonian $\hat{H}_{BP}$ of Eq.~(\ref{eq:mapHam}) contains a vacuum energy $E_{0}+n_{0}g_{IB}$ which we choose as the zero of the energy scale. Accordingly, 
$\hat{H}_{BP}|\mathrm{vac}\rangle=0$. We define $\{|\nu(\mathbf{k})\rangle \}$ as those eigenfunctions of $\hat{H}_{BP}$ with energy eigenvalue $E_{\nu}(\mathbf{k})$ and with one impurity with
momentum $\mathbf{k}$. Inserting a complete set of
eigenstates in Eq.~(\ref{eq:Greenstime}) gives
\begin{equation}
G(\mathbf{k},\tau)=-\theta(\tau)\sum_{\nu}|\langle \nu(\mathbf{k})|\hat
{c}_{\mathbf{k}}^{\dagger}|\mathrm{vac}\rangle|^{2}e^{-E_{\nu}(\mathbf{k}%
)\tau} \; .
\end{equation}
Under the conditions that the polaron is a stable quasi-particle in
the ground state (in the sense that it appears as a $\delta $-function
peak in the spectral function), one can extract its energy
$E_p(\mathbf{k})$ and $Z$-factor $Z_{0}$ by studying the long
imaginary time behavior of the polaron's Green's function:
\begin{equation}
G(\mathbf{k},\tau)\overset{\tau \rightarrow+\infty}{\sim}-Z_{0}(\mathbf{k}%
)~e^{-(E_p(\mathbf{k})-\mu)\tau},
\label{Greens}
\end{equation}
where the parameter $\mu$ is introduced to render a descending
exponential tail and
\begin{equation}
Z_{0}(\mathbf{k})=|\langle \Psi(\mathbf{k})|\hat{c}_{\mathbf{k}}^{\dagger
}|\mathrm{vac}\rangle|^{2},
\end{equation}
with $\Psi(\mathbf{k})$ the fully interacting ground state. The
asymptotic behavior of \cref{Greens} is associated with a pole
singularity for the Green's function in imaginary-frequency
representation. For $(E_p(\mathbf{k})-\mu)>0$ one has
\begin{equation}
\begin{split}
G(\mathbf{k},\omega)  & =\int_{0}^{+\infty}d\tau e^{i\omega \tau}%
G(\mathbf{k},\tau) \\
& =\frac{Z_{0}(\mathbf{k})}{i\omega+\mu-E_{p}(\mathbf{k})}
+\mathrm{~regular~part}\;.
\end{split}
\label{eq:Greensfreq}
\end{equation}
The one-body self-energy $\Sigma(\mathbf{k},\omega)$ is related to the
Green's function by means of the Dyson equation
\begin{equation}
 G(\mathbf{k},\omega)=\frac{1}{\frac{1}{G^0(\mathbf{k},\omega)}-\Sigma(\mathbf{k},\omega)} \; ,
\label{eq:dyson}
 \end{equation}
 with $G^0(\mathbf{k},\omega)$ the free impurity Green's function.  Since the Eqs.~(\ref{eq:Greensfreq}) and (\ref{eq:dyson}) possess the same
pole structure, the following expression for the
polaronic ground-state energy $E_{p}=E_{p}(\mathbf{k}=\mathbf{0})$ can be obtained \cite{Prokofev1998}:
\begin{equation}
E_{p}=\int_{0}^{\infty}d\tau \Sigma(\tau)e^{(E_{p}-\mu)\tau
}\;,\label{eq:Epolcalc}
\end{equation}
with $\Sigma(\tau)=\Sigma(\mathbf{0},\tau)$.   Calculating the Green's function boils down to summing a
series of Feynman diagrams over all topologies and orders, thereby
integrating over all internal variables (like momentum and imaginary
time). It is shown in \cite{Prokofev1998} that DiagMC is very suitable
to accurately compute the Green's function through a series expansion.
We consider irreducible diagrams (an example is shown in
Fig.~\ref{fig:irrdiag}) and evaluate a large number of diagrams $D$ in
order to numerically compute the $\Sigma(\mathbf{p},\tau)$
\begin{equation}%
\begin{split}
\Sigma & (\mathbf{p},\tau) =   \sum_{n=0}^{\infty}\sum_{\xi_{n}} \sum_{\mathbf{q_{i=1,\ldots,n}}} \int d\tau_{1} \ldots
d\tau_{i} \ldots d\tau_{n} \\
& \times D(\xi_{n},\mathbf{p},\tau,\tau_{1},\ldots,\tau_{i},\ldots,\tau
_{n},\mathbf{q_{1}},\ldots,\mathbf{q_{i}},\dots,\mathbf{q_{n}}) \; ,
\end{split}
\label{Sigma}
\end{equation}
where $\xi_{n}$ represents the topology, $n$ the diagram order, $\mathbf{q_{i}%
}$ is the internal momentum and  $\tau_{i}$ is the imaginary time. The DiagMC
technique allows one to sample over all topologies, all orders and all values
of the internal variables. 

In \cref{fig:irrdiag} some Feynman diagrams for $\Sigma(\tau)$ are
shown. The algebraic expression for these diagrams is given in terms
of free propagators and interaction vertices:
\begin{itemize}
\item[(i)] The free-impurity propagator in imaginary time is determined by 
\begin{equation}
G^{(0)}(\mathbf{k},\tau)=-\theta(\tau) e^{-(\epsilon
_{k}-\mu)\tau}\;.
\end{equation}
\item[(ii)] The propagator for an elementary phonon excitation, either
  of the Bogoliubov type for the BEC polaron, or acoustic phonons for
  the acoustic polaron is defined in Eq.~(\ref{eq:phononpropagator}).
\item[(iii)] A vertex factor $V(\mathbf{q})$ whenever an elementary excitation carrying momentum
$\mathbf{q}$ is created or annihilated.
\end{itemize}
The diagram order is defined by the number of elementary
excitations. 
\begin{figure}[h]
\includegraphics[width=\columnwidth] {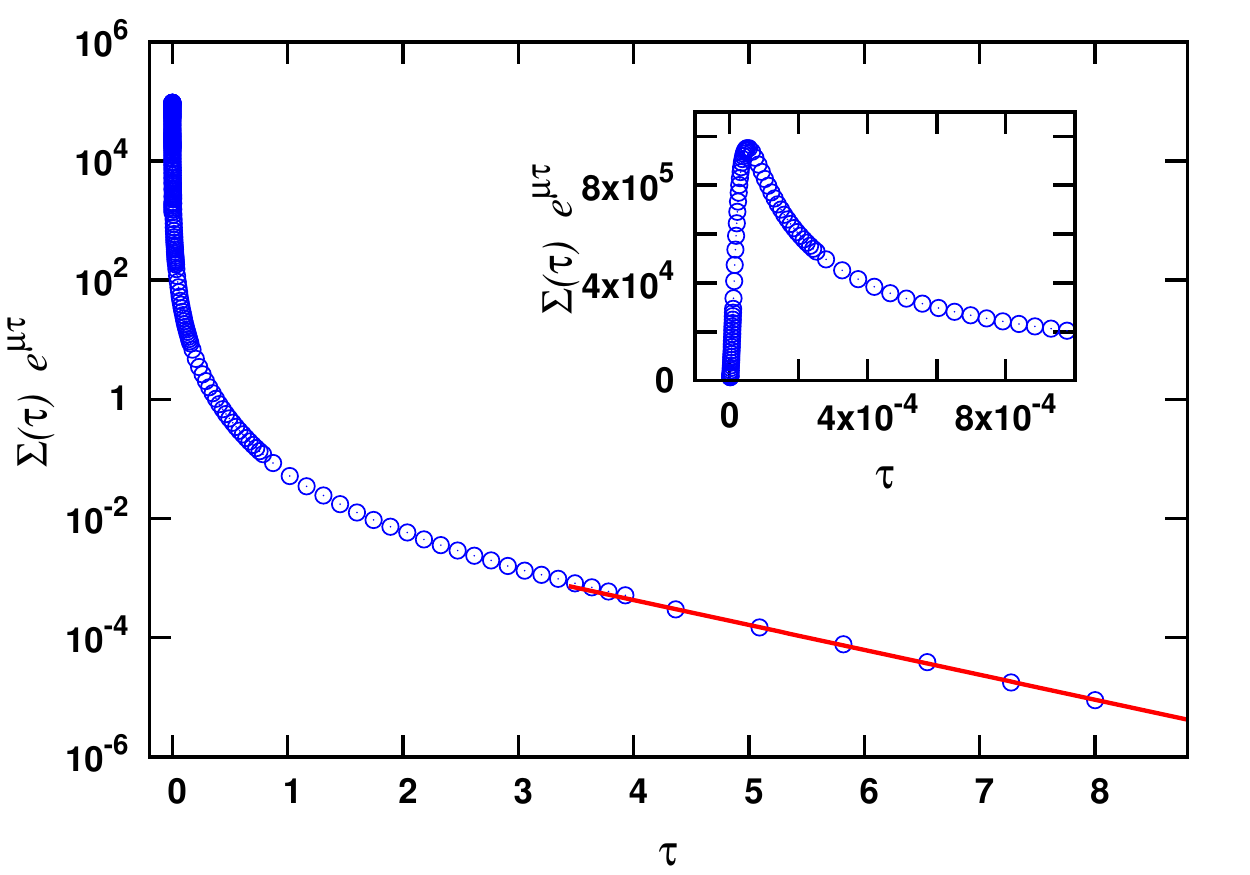}
\caption{(Color online) The one-body self-energy $\Sigma(\tau)
  e^{\mu\tau}$ for $\mu=-2$ for the BEC polaron plotted as a function of imaginary
  time $\tau$.  Results are obtained for $\alpha_{IB}=5$ and $q_c=200$
  and exclude the first-order contribution to $\Sigma(\tau)
  e^{\mu\tau}$ which can be easily computed analytically. The inset shows
  $\Sigma(\tau) e^{\mu\tau}$ for small imaginary times.  }
\label{fig:sigma} 
\end{figure}

\section{Results and discussion}\label{result}

\subsection{BEC polaron}\label{becresult}
For the Fr\"{o}hlich polaron for which the ground-state energies are
displayed in Fig.~\ref{fig:EpolEl}, the one-body self-energy
$\Sigma(\tau)$ can be computed by means of the procedure sketched in
Sec.~\ref{diagmc}.  For the BEC polaron, on the other hand, one
encounters ultraviolet divergences when evaluating $\Sigma(\tau)$ and
its energy cannot be extracted.  Renormalization/regularization of the
impurity-boson pseudopotential is required to obtain physically
relevant results for the energies. As a first step in the
renormalization procedure, we introduce a momentum cutoff $q_c$ upon
replacing the momentum summations in Eq.~(\ref{eq:mapHam}) by
integrals:
\begin{equation}
  \sum_{\mathbf{k}} \rightarrow \frac{\mathcal{V}}{(2\pi)^3} 
\int_{|\mathbf{k}|<{q_c}} d\mathbf{k}\;.
\end{equation}
This allows us to calculate $\Sigma(\tau)$ and the accompanying
ground-state energy $E^{MC}_p$. From now on we will make the
distinction between the polaron energy calculated by DiagMC
($E^{MC}_p$) and calculated by the Feynman approach
($E^F_p$). Obviously, $E^{MC,F}_p$ depends on $q_c$ and in order to
stress this dependence we use the notation $E^{MC,F}_p(q_c)$.  In
\cref{fig:sigma} we show an example of the time dependence of the
one-body self-energy $\Sigma (\tau)$ for the BEC polaron for
$q_c=200$. As can be noticed, after introducing a momentum cutoff
$q_c$, the $\tau$ dependence is well behaved and the asymptotic regime
of $\Sigma (\tau)$ can be identified.  The $\sum_{n=0}^{\infty}$ in
\cref{Sigma} implies a summation over an infinite number of diagram
orders. In practice, we set a cutoff $N_{\text{max}}$ for $n$ in
evaluating $\Sigma(\tau)$. For each $N_{\text{max}}$ we can find a
corresponding imaginary time $\tau_{\text{max}}$ for which
$\Sigma(\tau)$ converges. Upon increasing $N_{\text{max}}$ we can
choose a larger value for $\tau_{\text{max}}$. An optimal
$N_{\text{max}}$ is reached when we can find a $\tau_{\text{max}}$ in
the asymtotic regime that allows us to fit the tail of
$\Sigma(\tau)$. In this way we make an extrapolation for $\tau
\rightarrow \infty$ which determines the value $N_{\text{max}}$.
Typical values of $N_{\text{max}}$ are of the order $10^4$ for large
values of $\alpha_{IB}$. With the aid of the Eq.~(\ref{eq:Epolcalc}),
$E^{MC}_p(q_c)$ can be extracted from the computed $\Sigma
(\tau)$. The error on $E^{MC}_p(q_c)$ contains a statistical error and
a systematic error stemming from the fitting procedure.  As can be
appreciated from \cref{fig:sigma}, the grid in imaginary time has to
be chosen carefully, since the short-time behavior of $\Sigma (\tau)$
is strongly peaked. The $\Sigma(p,\tau)$ for these short times
delivers a large contribution to the energy.
\begin{figure}[h]
\includegraphics[width=\columnwidth] {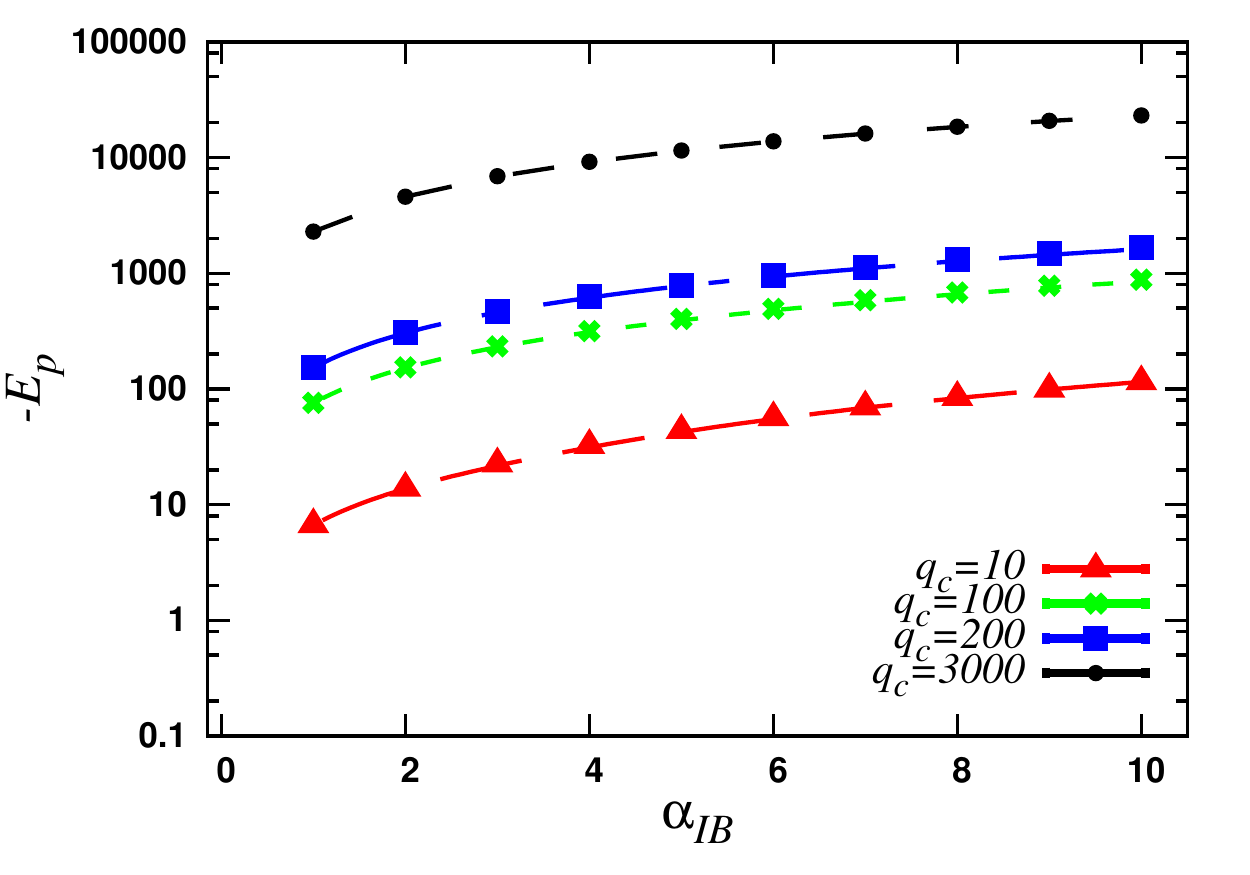}
  \caption{(Color online) The non-renormalized BEC-polaron energy
    $E_p$ as a function of the coupling strength $\alpha_{IB}$ as
    computed with the DiagMC (symbols) and with the Feynman (lines)
    approaches. Results are shown for four values of the cutoff
    momentum. }
\label{fig:EpolBecwithout} 
\end{figure}
\par In \cref{fig:EpolBecwithout}, results for the non-renormalized
energies $E^F_p(q_c)$ and $E_p^{\text{MC}}(q_c)$ are presented as a
function of the dimensionless coupling parameter $\alpha_{IB}$ defined
in \cref{AlfBEC}. The $\alpha_{IB}$ and $q_c$ dependence of the DiagMC
energies is remarkably similar to those of the Feynman energies.  We
observe that $E^{\text{MC}}_p(q_c)$ lies a few percent below
$E^F_p(q_c)$ for all combinations of $\alpha_{IB}$ and $q_c$
considered.  \par
In Ref.~\cite{PhysRevB.80.184504} a renormalization procedure to
eliminate the $q_c$ dependence of the computed polaron energy is
outlined. When determinig the $T$-matrix of Eq.~(\ref{eq:lad}) up to
second order, the following relation between the scattering length
$a_{IB}$ and the coupling strength $g_{IB}$ is obtained:
\begin{equation}
 \frac{2\pi a_{IB}\hbar^2}{m_r}=g_{IB}-\frac{g_{IB}^2}{(2\pi)^3} 
\int_{|\mathbf{q}|<q_c} d\mathbf{q} \frac{2m_r}{\hbar^2q^2}  \; .
\end{equation}
Using this expression, the $n_0g_{IB}$ term in \cref{eq:mapHam} can be
replaced by :
\begin{equation}
 n_0 g_{IB} \rightarrow  \frac{2\pi a_{IB}n_0\hbar^2}{m_r} +E_{\text{ren}}(q_c) \; ,
\end{equation}
whereby we have defined $E_{\text{ren}}(q_c)$ :
\begin{equation}
 E_{\text{ren}}(q_c)= \frac{n_0 g_{IB}^2}{(2\pi)^3}
\int_{|\mathbf{q}|<{q_c}} d\mathbf{q} \frac{2m_r}{\hbar^2 q^2} \;. 
\label{eq:LS}
\end{equation}
This renormalization procedure was developed in the context of the
Feynman approach \cite{PhysRevB.80.184504}. The same procedure can 
also be applied in the DiagMC framework.  In both frameworks, the
renormalized polaron ground-state energy can be found by evaluating the sum
\begin{equation}
E^{MC,F}_p=E_p^{MC,F}(q_c \rightarrow \infty)+E_{\text{ren}}(q_c
\rightarrow \infty) \; .
\label{eq:energyrenorma}
\end{equation}

  \begin{figure}[h]
  \includegraphics[width=\columnwidth] {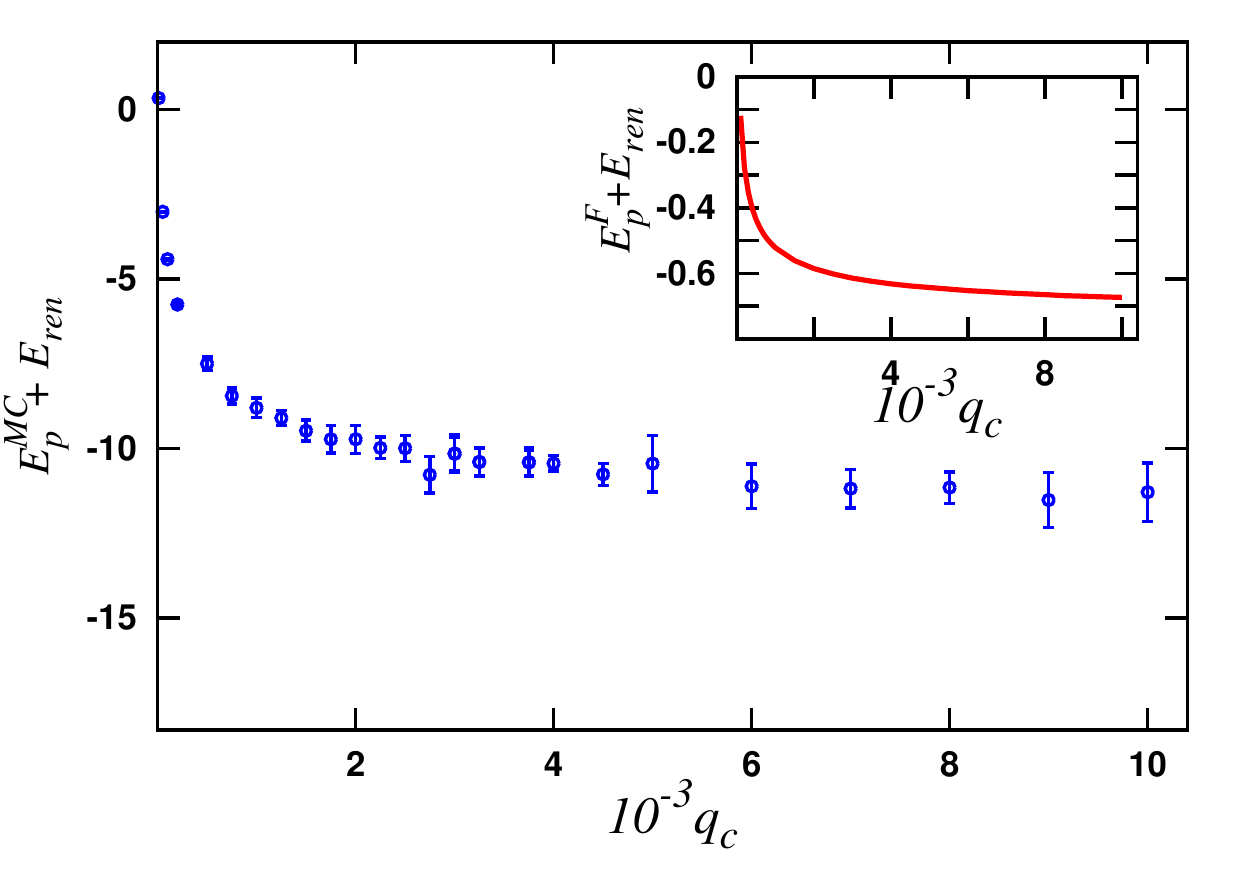}
  \caption{(Color online) The renormalized BEC-polaron energies
    $[E_p^{MC}(q_c)+E_{\text{ren}}(q_c)]$ at
    $\alpha_{IB}=3$ are given as a function of the momentum cutoff
    $q_c$. The inset figure shows $[E_p^{F}(q_c)+E_{\text{ren}}(q_c)]$
    as a function of $q_c$.  }
\label{fig:EpolBec_conv} 
  \end{figure}
\begin{figure}[h]
  \includegraphics[width=\columnwidth] {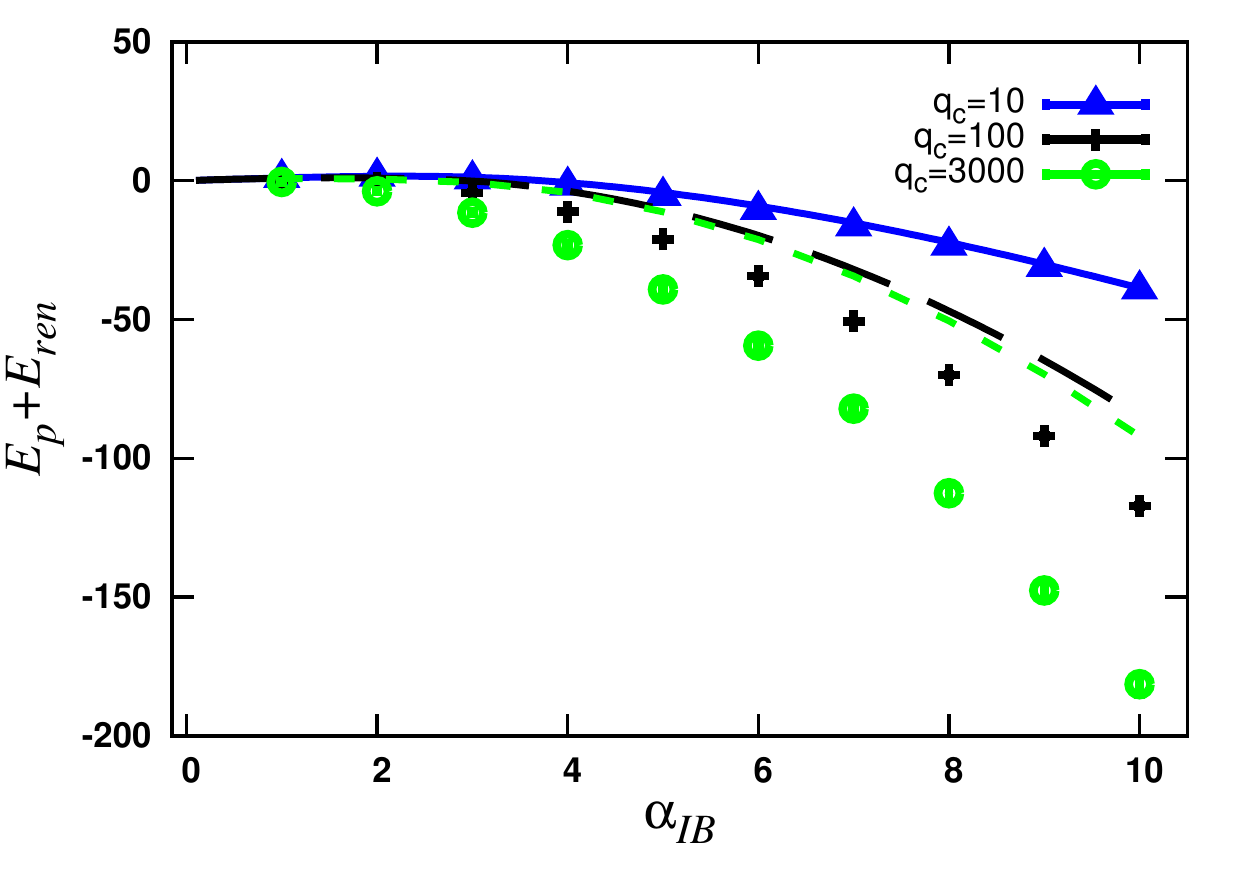}
  \caption{(Color online) The renormalized BEC-polaron energies
    $[E_p(q_c)+E_{\text{ren}}(q_c)]$ as a function of
    $\alpha_{IB}$ for different values of the momentum cutoff
    $q_c$. Lines are the Feynman path-integral and symbols are the
DiagMC results.}
\label{fig:EpolBEC} 
  \end{figure}
\begin{figure}[h]
  \includegraphics[width=\columnwidth] {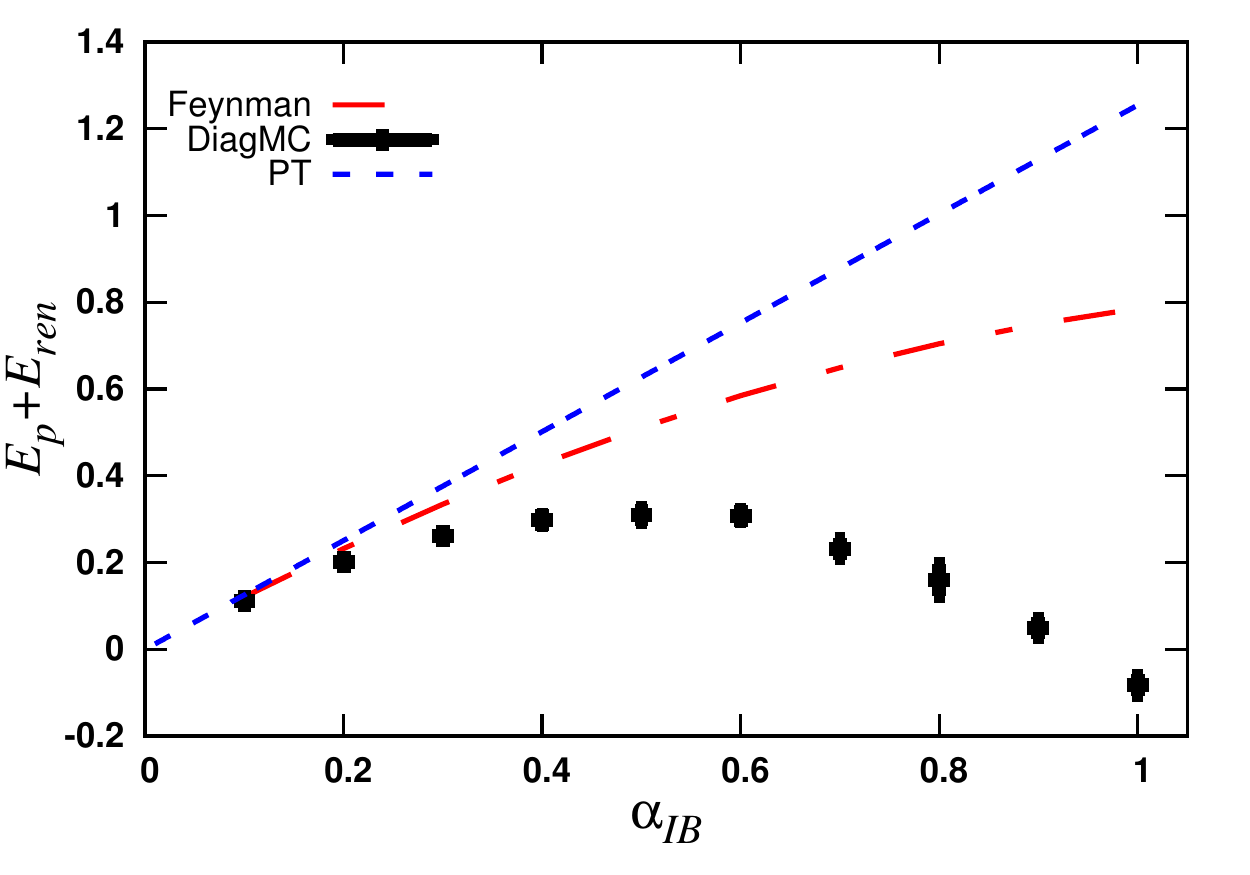}
  \caption{(Color online) The renormalized BEC-polaron energies
    $[E_p(q_c)+E_{\text{ren}}(q_c)]$ at small values of $\alpha_{IB}$
    at $q_c=2000$. The dot-dashed line is the Feynman path-integral
    result, symbols represent the DiagMC results, while the short
    dashed line is the prediction from second-order perturbation
    theory (PT).}
\label{fig:EpolBECPert} 
  \end{figure}

  In order to illustrate the convergence of the
  Eq.~(\ref{eq:energyrenorma}) in both approaches, in
  \cref{fig:EpolBec_conv} the energies
  $[E_p^{MC}(q_c)+E_{\text{ren}}(q_c)]$ and
  $[E_p^{F}(q_c)+E_{\text{ren}}(q_c)]$ are plotted as a function of
  $q_c$ for a representative value $\alpha_{IB}=3$ of the coupling
  strength. We notice that the DiagMC and the Feynman approach display
  an analogous $q_c$ dependence. Convergence is reached for $q_c
  \gtrsim 3000$.  Fig.~\ref{fig:EpolBEC} shows that 
the Feynman path-integral predictions for the BEC-polaron ground-state
energies overshoot the DiagMC ones. The relative difference between
the two predictions increases with growing values of $q_c$.  The very
good agreement between the two methods that was found in
\cref{fig:EpolBecwithout} for the non-renormalized energies, is no
longer observed for the renormalized energies. Indeed, the latter are
obtained with \cref{eq:energyrenorma}, which amounts to substracting
two numbers of almost equal magnitude.  Accordingly, the final result
for the renormalized BEC-polaron ground-state energy is highly
sensitive to the adopted many-body technique and renormalization
procedure.
Fig.~\ref{fig:EpolBECPert} illustrates
that for small $\alpha_{IB}$ both methods reproduce the result from
second-order perturbation theory.

\par
\begin{figure}[h]
 \includegraphics[width=\columnwidth] {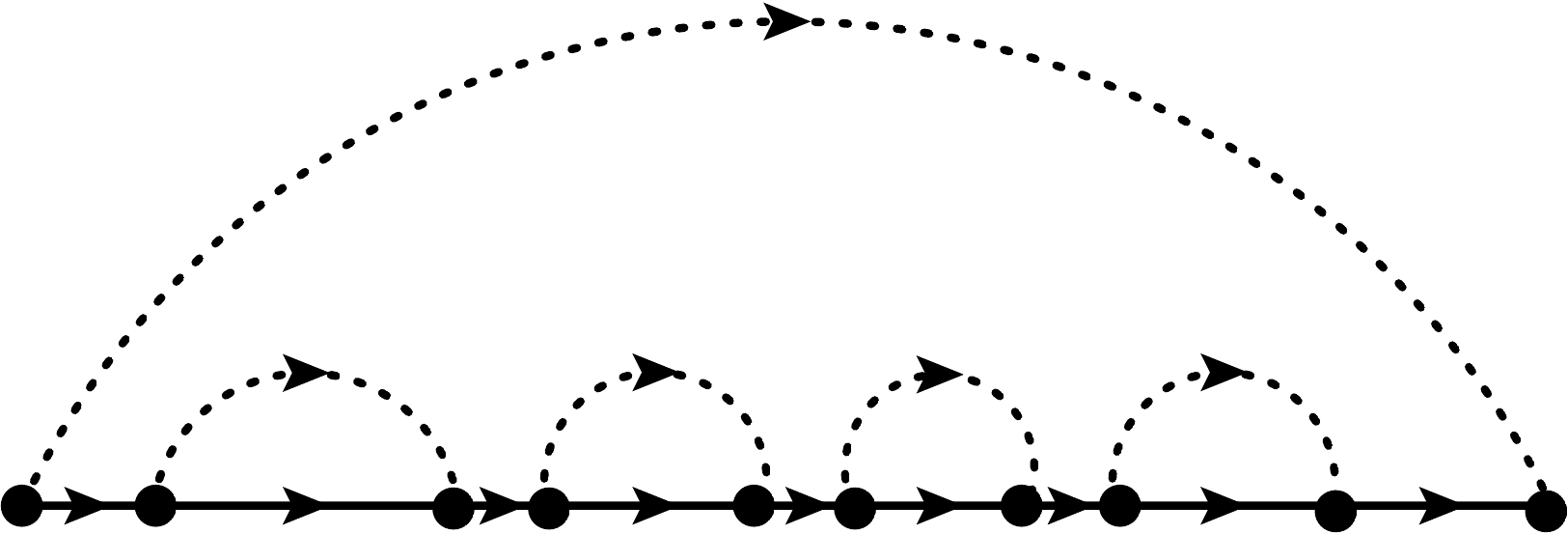}
\caption{A diagram of order five for the one-body self-energy. Line
  conventions as in Fig. \ref{fig:irrdiag}. Imaginary time runs from
  left to right. A first-order subdiagram occurs whenever a
  first-order diagram drops out from the full diagram by cutting the
  solid line at two selected times. For example, the considered
  diagram contains four first-order subdiagrams.}
\label{fig:domindiag} 
  \end{figure}

 The DiagMC method samples diagrams according to their weight and
it can be recorded how many times a specific diagram is sampled. In
this way, one can identify those diagrams with the largest weight in
the self-energy $\Sigma(\tau)$.
At fixed diagram order, we have observed that the number of
first-order subdiagrams--the definition of which is explained in the
caption of \cref{fig:domindiag}--plays a crucial role in the weight of
the diagram. Our studies indicate that for $q_c>50$ the most important
diagram is the one with the highest number of first-order
subdiagrams. We have considered many combinations of $\alpha_{IB}$ and
$q_c$ and could draw this conclusions in all those situations. The
dominance of this diagram becomes more explicit with increasing values
of $q_c$.

\subsection{Acoustic polaron}\label{acresult}
\begin{figure}[ht]
  \includegraphics[width=\columnwidth] {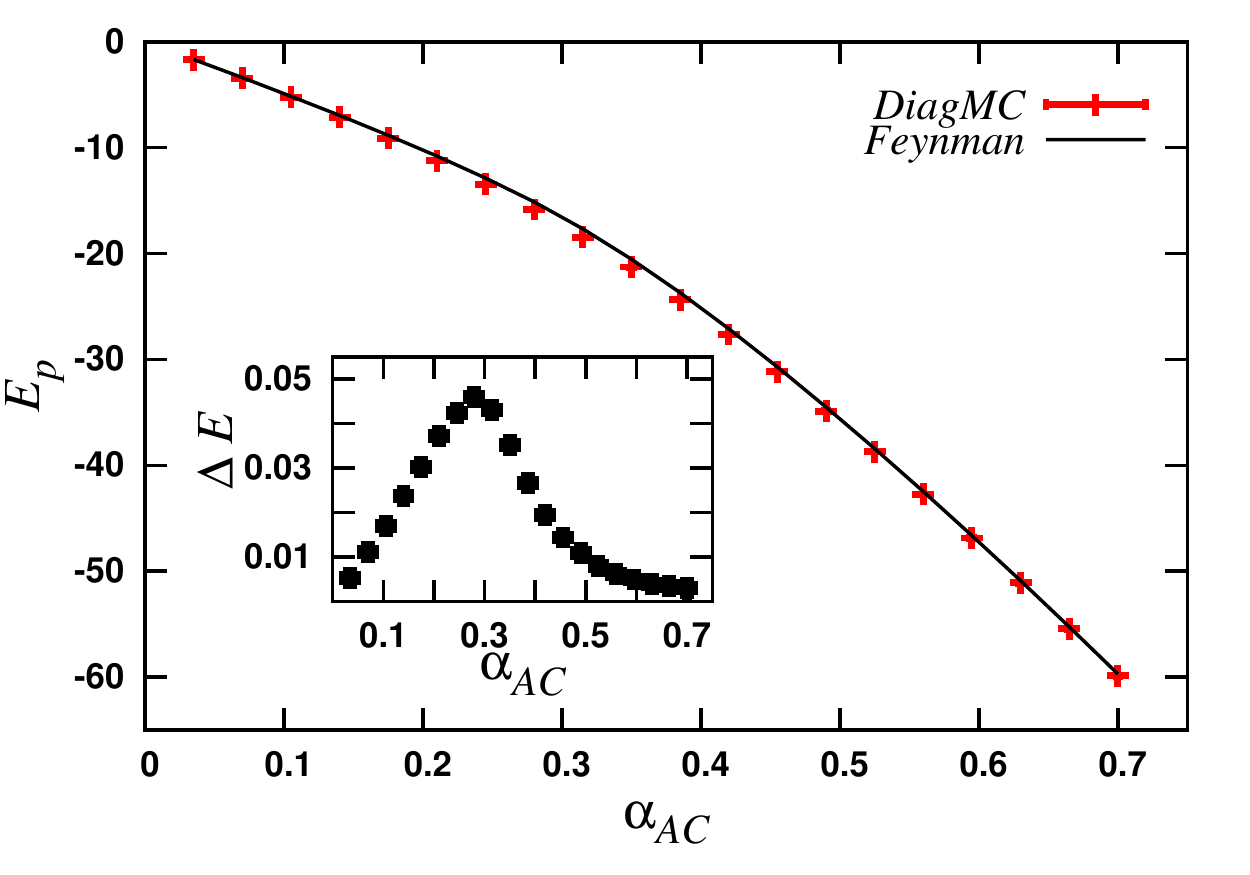}
  \caption{Non-renormalized ground-state energies $E_p^F$ and $E_p^{MC}$ for the
    acoustic polaron as a function of $\alpha_{AC}$ for $k_0=10$. The
    inset shows $\Delta E=\frac{E^{MC}_p-E^F_{p}}{E^{MC}_p}$ as a
    function of $\alpha _{AC}$.}
  \label{fig:Acrel10} 
\end{figure}

\begin{figure}[ht]
  \includegraphics[width=\columnwidth] {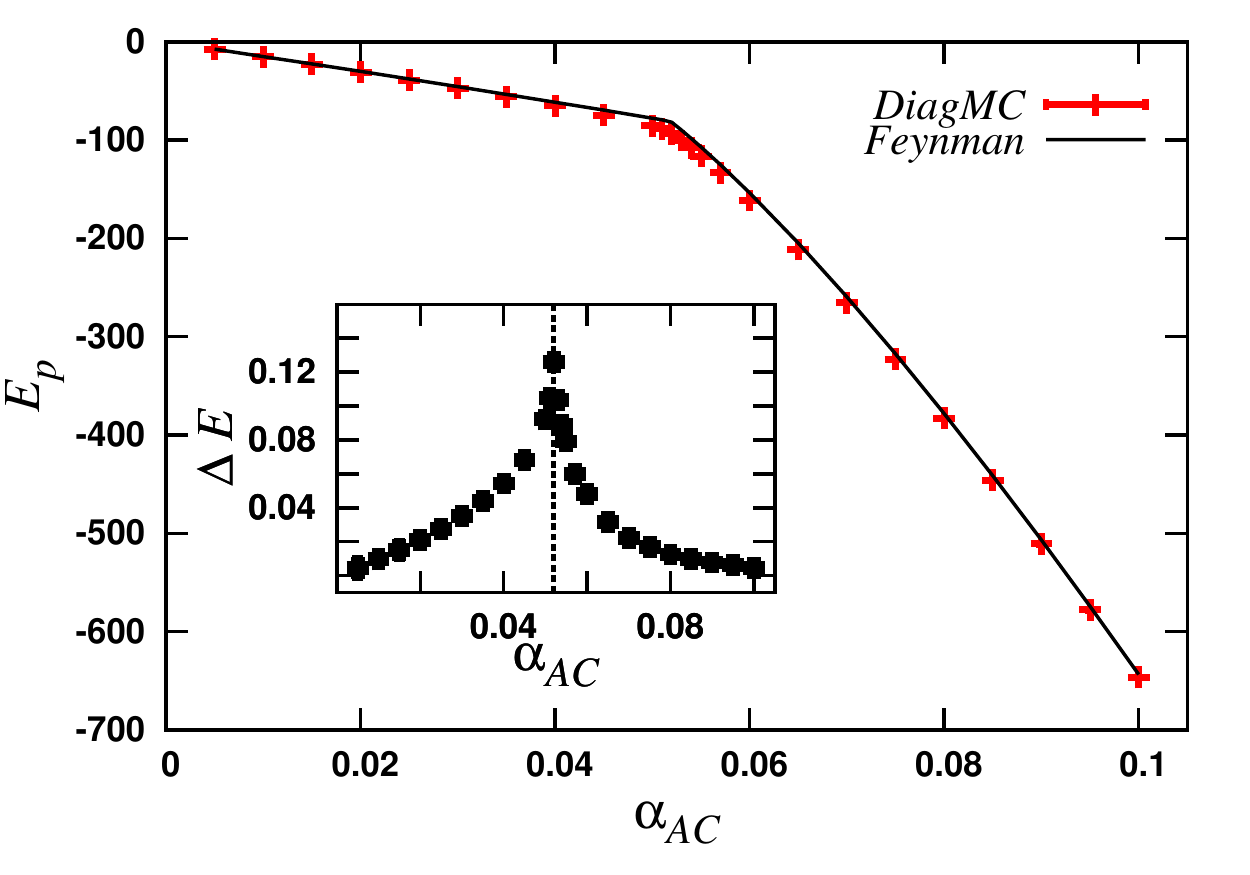}
  \caption{As in Fig.~\ref{fig:Acrel10} but for $k_0=50$. The vertical
    dashed line denotes the coupling strength $\alpha_{AC}=0.052$
    corresponding with the transition as computed in
    Ref.~\cite{PhysRevB.32.3515}.}
  \label{fig:Acrel50} 
\end{figure}

We now discuss the numerical results for the ground-state energy of
the acoustic polaron introduced in Sec.~\ref{ac}.  In
Figs.~\ref{fig:Acrel10} and \ref{fig:Acrel50} we show a selection of the
predictions $E^F_p$ from the Feynman upper-bound method of
Eq.~(\ref{UppBound}) together with the DiagMC results $E^{MC}_p$ which
are computed with the aid of Eq.~(\ref{eq:Epolcalc}). For $k_0=10$ and
$k_0=50$ an excellent agreement between $E^F_p$ and
$E^{MC}_p$ is found. From the relative difference $\Delta
E=\frac{E^{MC}_p-E^F_{p}}{E^{MC}_p}$, a value $\alpha_{AC}$ can be
found where $\Delta E$ is largest in the considered region of
$\alpha_{AC}$. For $k_0=10$ we find $\alpha_{AC}^{k_0=10}=0.28 \pm
0.04$ and for $k_0=50$, $\alpha_{AC}^{k_0=50}=0.52 \pm 0.01$. For 
$\alpha < \alpha_c$, $\Delta E$ increases with $\alpha_{AC}$ and for 
$\alpha > \alpha_c$ $\Delta E$ decreases with increasing $\alpha_{AC}$.  We remark that $\alpha_c^{k_0=10}$ and $\alpha_c^{k_0=50}$
coincides with the coupling strength for the transition
\cite{doi:10.1143/JPSJ.35.137} as computed with the Feynman
approach. \par
From a detailed analysis of the DiagMC results for $k_0=50$ we find
that the class of diagrams of the type sketched in
\cref{fig:domindiag} plays a dominant role for $\alpha _{AC} <
\alpha_{c}$. For $\alpha _{AC} > \alpha_{c}$ we observe a dramatic
change in the importance of those diagrams, and we can no longer
identify a class of a diagrams that provides the major contribution to
the self-energy $\Sigma(\tau)$. \par The knowledge of a certain class
of dominant diagrams can be exploited to develop approximate schemes.
Indeed, one can set up a self-consistent scheme thereby summing over
an important class of diagrams, including the observed dominant
ones. In practice, the procedure can be realized by introducing bold
(or dressed) propagators
\begin{equation}
\begin{split}
  \Sigma^{(i-1)}(\mathbf{p},\omega) & = \int d\omega ^{\prime} \int
\frac{d\mathbf{q}}{(2\pi)^3}  \\
& \times G^{(i-1)}(\mathbf{p}-\mathbf{q},\omega-\omega^{\prime}) \mathcal{D}(\mathbf{q},\omega ^{\prime})  \\
G^{(i)}(\mathbf{p},\omega) & = \frac{1}{{G^0}^{-1}(\mathbf{p},\omega)-\Sigma^{(i-1)}(\mathbf{p},\omega)} \; \; , 
\end{split}
\end{equation}
with $\omega$ and $\omega^{\prime}$ the imaginary frequencies.  
The self-energy $\Sigma^{(i-1)}$ and the dressed Green's function
$G^{(i)}(\mathbf{p},\omega)$ are calculated for subsequent values of
$i$, starting from $i=1$, until $G^{(i)}(\mathbf{p},\omega)$ is
converged. In this way $ \Sigma^{(i)}(\mathbf{p},\omega)$ will contain
all diagrams for which the lines of the phonon propagators do not
cross.

 \FloatBarrier
\section{Conclusions}\label{concl}
We have studied the ground-state energies of the BEC polaron and the
acoustic polaron, two large polaron systems that can be described by a
Fr\"{o}hlich type of Hamiltonian. When calculating energies for the
BEC polaron with the DiagMC and the Feynman variational technique, we
encounter similar ultraviolet divergences. For the acoustic polaron, the
ultraviolet regularization is achieved by a hard momentum cutoff which
is naturally set at the edge of the first Brillouin zone. In this
case, the DiagMC and Feynman predictions for the ground-state energies
agree within a few percent. The largest deviation between the
predictions of both methods, was found at a coupling strength that
marks the transition between a quasifree and a self-trapped state. For
the BEC polaron, a more involving two-step renormalization procedure
is required. The first step is the introduction of a hard momentum
cutoff. In line with the results for the acoustic polaron, the DiagMC
and Feynman non-renormalized ground-state energies of the BEC polaron which are
produced in this step are remarkably similar. Therefore, one can infer
that the Feynman variational method reproduces the ``exact'' DiagMC
non-renormalized polaron ground-state energies at a finite momentum
cutoff.

In order to obtain the physical, or renormalized, BEC-polaron energies
from the non-renormalized ones, an additional procedure is required.
Thereby, the contact interaction is renormalized with the aid of the
lowest-order correction obtained from the Lippmann-Schwinger
equation~(\ref{eq:LS}). Despite the fact that the absolute difference
between the Feynman and DiagMC BEC-polaron energies remains unaffected
by this procedure, the final result for the physical energies displays
a large discrepancy.

\acknowledgments This work is supported by the Flemish Research
Foundation (FWO Vlaanderen) through project numbers G.0119.12N and G.0115.12N. Discussions with S.N.~Klimin and
L.A.~Pena-Ardila are gratefully acknowledged. The
computational resources (Stevin
Supercomputer Infrastructure) and services used in this work were
provided by Ghent University, the Hercules Foundation, and the Flemish
Government.

\end{document}